# Soei: Structured Observability of Evidence and Intent for Multi-Agent AI Software Engineering

*An empirical study of 45 real coding-agent sessions*
*and a provenance-grounded architecture for verifiable agent work*

Mohammed Nayeem Khan
mohammed.nayeem.khan@intel.com
mnknayeem@gmail.com

Sarun Kumar K S
sarun.kumar.k.s@intel.com

Intel Corporation

**Abstract**

AI coding agents now perform long, multi-step and multi-session software engineering work with limited human supervision. When a developer returns to such work, the decisive question is not "what did the model emit?" but "what was actually done, by whom, and can I trust that it is finished?". We study this question in two ways. **First**, we build a working prototype of our proposed system and run it on a real corpus of **45 coding-agent sessions** (102,233 raw records; 20,577 tool executions; 2,336 human-to-agent episodes; April–October 2026). With a blindly audited classifier (precision 83.3% / 96.7%), we estimate that **23.4% [95% CI 17.1–35.7] of code-editing episodes end without any independent verification** after the last edit, and that **57.8% of to-dos the agent marked complete after editing had no proof** between edit and completion. A 10-minute timing heuristic for session hand-offs recovers only **6 of 24 (25%)** session pairs that our content-lineage analysis links, because **95.5%** of real cross-session artifact reuse happens more than ten minutes apart. 42% of sessions span more than 24 hours, and 48.1% of transcript bytes lie beyond a 2 MiB replay bound. Credential-like strings appear in 1,326 raw records across 16 of 45 sessions. **Second**, we propose **Soei** (Structured Observability of Evidence and Intent): an acknowledgement-aware event schema, a hash-chained local evidence ledger, a W3C-PROV-aligned evidence graph, calibrated inference for attribution, drift and lineage-based hand-offs, and a verification layer that separates an agent's *claim* of completion from *proof*. On the real corpus, the prototype's ledger is 20% of the raw transcript size, stores zero content and zero credential matches, verifies at about 126k events/s, detects tampering, and replays deterministically. We formalise the model, specify the architecture in implementation detail, and define a reproducible evaluation agenda. Finally, we argue that this evidence substrate is a prerequisite for a **golden harness** for evaluating and certifying coding agents, and for the **governance of autonomous agents**. We define autonomy levels with evidence gates and harness metrics whose baselines come from our data.

**Key empirical findings (real data, this paper)**

1. **The claim–proof gap is real:** about one in four code-editing episodes ends unverified (23.4%, CI 17.1–35.7%). Across 4 classifier versions the raw rate stays between 25.1% and 42.1%.
2. **"Done" is not proven:** 186 of 322 (57.8%) to-dos completed after edits had no verification between the edit and the "completed" status.
3. **Hand-offs are slow, not fast:** real cross-session reuse has a median gap of 8.8 days. A ≤10 min temporal heuristic finds 25% of linked pairs; content lineage finds all 24.
4. **Bounded history loses most of the evidence:** 19/45 sessions exceed 24 h, and 48.1% of bytes fall beyond a 2 MiB replay window.
5. **Acknowledgement matters:** 40 of 3,548 file mutations (1.1%, in 30 episodes) were requested but never succeeded. A start-event observer would show them as applied.
6. **Raw agent logs are a liability:** 1,326 records in 16/45 sessions contain credential-like strings (208 high-confidence token matches, 5 distinct tokens). The Soei ledger contains none.
7. **Provenance is cheap:** the hash-chained ledger costs 20% of raw size and verifies at about 126k events/s in pure Python, with deterministic replay and tamper detection.

## 1 Introduction

Coding agents have progressed from single-turn completion to long-horizon work. They read repositories, plan, edit many files, run tools and tests, and increasingly operate as several sessions or sub-agents. Issue-resolution benchmarks such as SWE-bench [1], agent–computer interfaces such as SWE-agent [2] and reason-act tool use [3] have driven rapid capability gains. The *supervisory* side of human–agent collaboration has received far less engineering and scientific attention. A developer returning to a running agent must rebuild state from scrolling transcripts across terminals. This maps directly onto the three levels of situation awareness (perception, comprehension, projection) [4].

Two families of tools address the problem in part. *LLM trace dashboards* (LangSmith, Langfuse, Arize Phoenix) and the OpenTelemetry generative-AI semantic conventions [5] record model calls, prompts, latency and token cost. They answer "what did the model do?" but not "which part of my project moved, and is it really done?". *Project-level observability tools* watch the repository, a plan file and whatever agent events a vendor exposes, and visualise progress at the level of components and tasks. The better ones run locally and stay read-only. None of them, as far as we can find, measures how far what they show is from what actually happened.

We argue that visibility alone does not give a supervisor what they need. The evidence behind each view has to be trustworthy, and that is a testable claim. We therefore (i) measure the evidence problems that observability must solve on a real corpus of coding-agent sessions, and (ii) design, prototype and evaluate Soei, an architecture that treats agent reports as *claims*, stores *evidence before inference*, and derives every view from one tamper-evident record. Several 2026 papers reach related conclusions from other directions (§2.6); ours is, to our knowledge, the first to measure the claim–proof gap and the hand-off recall problem on real working sessions and to build the substrate as a passive observer of unmodified vendor agents.

**Contributions.**

1. **A gap taxonomy** of 14 gaps across six dimensions, each mapped to a requirement (§3, App. B).
2. **The Soei core architecture**, specified to implementation level: event schema, acknowledgement tracker, privacy gate, ledger format, PROV-aligned graph, inference engines, claim–proof verifier, projection API, and the trust and deployment model (§5, §7–8).
3. **A formal model** with an evidence lattice, task state machine, confidence fusion, graded hand-offs, drift and completion trust (§6).
4. **An empirical pilot on real data:** 45 sessions, an audited claim–proof classifier with a bias-corrected estimate, lineage-versus-heuristic hand-off analysis, history-bound loss, acknowledgement failures, secret exposure and ledger cost (§9–10).
5. **Reproducible artifacts:** a stdlib-only prototype, frozen aggregate results, figure scripts and audit sheets (App. E).
6. **A path to a golden harness and agent governance:** autonomy levels with evidence gates, harness metrics with real-data baselines, a mapping to NIST AI RMF, ISO/IEC 42001 and the EU AI Act, and a prioritised improvement roadmap (§13–14).

## 2 Background and related work

### 2.1 Coding agents and their artifacts

Modern coding agents interleave reasoning and tool actions [3] through repository-specific interfaces [2] and are measured on issue resolution [1]. In practice they run inside vendor tools (VS Code Copilot agent, Codex CLI/desktop,

Claude Code, Cursor) that persist session logs and expose hooks and native to-do lists. These artifacts are the raw material for observability. They are also vendor-specific and incomplete, and their format changes without notice. As we show in §10.6, they may also contain secrets.

### 2.2 Tracing, clocks and observability

Distributed tracing (Dapper [6], later OpenTelemetry) models work as spans with causal parents and propagated context. LLM observability adapts this to model and tool calls, and the OpenTelemetry GenAI conventions [5] standardise model-invocation attributes. AgentTrace [23] instruments agents at runtime with structured security-oriented logs, and OpenRath [21] proposes a first-class *Session* runtime value that is branchable, inspectable and replayable, so that transcripts, tool effects and lineage stop being recorded in separate places. Both require the agent to be built on, or instrumented by, their framework. These approaches assume instrumented code paths and centre on requests rather than on *project intent* (plans, components, deliverables). Logical clocks [7] and hybrid logical clocks [11] address ordering across sources without synchronised time. That problem returns when provider logs, file-system events and CI results are fused.

### 2.3 Provenance

The W3C PROV model [8] represents *entities*, *activities* and *agents* with relations such as *wasGeneratedBy*, *used*, *wasAttributedTo* and *wasDerivedFrom*. Provenance is a natural basis for questions such as “which session produced this exact file version, and did another session consume it?”. We show empirically that these questions cannot be answered from timing alone.

### 2.4 Visualisation and situation awareness

The “overview first, zoom and filter, then details on demand” principle [9] and Endsley’s situation-awareness model [4] set principled targets for a supervisory display: perception of current activity, comprehension of progress relative to intent, and projection of what happens next. SeaView [30] visualises SWE-agent trajectories for post-hoc error analysis; it reads benchmark trajectories rather than live developer sessions and does not separate claims from evidence.

### 2.5 Positioning

**Table 1.** Positioning of observability approaches for AI coding agents.

| Capability | Transcripts | LLM trace dashboards | Soei |
|---|---|---|---|
| Unit of analysis | Message | Model/tool span | Claim, evidence, artifact@digest, task, component |
| Project intent | No | No | Yes, plus a quantified drift metric |
| Evidence semantics | Implicit | Implicit | Formal lattice + calibrated confidence |
| Completion | Agent says so | N/A | Verified only by proof |
| Hand-offs | No | Parent/child spans | Graded H0–H3, content lineage |
| History | Scrollback | Server DB | Append-only hash-chained ledger |
| Privacy | Raw | Usually cloud | Local; one ingest-time policy; zero content |
| Measured accuracy | — | — | Audited classifier, CIs, ablations |

### 2.6 Concurrent work (2026)

Several papers published while this work was in progress treat agent output as a claim that needs independent evidence, and we position Soei against them here. **Proof-or-Stop** [18] gates lifecycle transitions (reviewed, tested, done, ready-to-merge) on fresh, source-state-bound, mechanically checkable evidence, and the **runtime-contract** position paper [19] argues for an “evidential face” of agent safety that gates task submission on test runs, log captures and diffs. Both *enforce*. Soei’s core is read-only by design (P1) and its contribution is to measure how large the gap is in ordinary use and to build the evidence substrate that such gates would consume; the policy plane we propose in §13.2 and §14 is where enforcement would live. **Artifact-centred claim-aware observability** [20] makes the same claim/artifact/relation argument for autonomous scientific agents; we make it for software work, add acknowledgement and hand-off lineage, and quantify it. **Skill Runtime Intelligence** [24] passively reconstructs Skill-lifecycle stages across heterogeneous coding-agent harnesses with four evidence grades and keeps unsupported stages as *unknown*, which is the closest match to our evidence lattice and adapter stance; it targets Skill activation rather than project-level claims, hand-offs or history integrity. **Auditable Agents** [22] defines auditability dimensions (action recoverability and others) that our four governance questions in §13.1 map onto; the hash-chained ledger and deterministic replay are one concrete way to satisfy them. On the empirical side, **AgentLens** [25] finds that 10.7% of *passing* OpenHands trajectories on SWE-bench Verified show regression cycles, blind retries or missing verification, which corroborates Finding 1 from a benchmark rather than a workplace corpus, and a large study of agent Skills [26] finds credential leakage through privileged skill code, a different channel from the transcript leakage we measure in §10.6. **TraceTrial-Bench** [27] and **HarnessAudit** [28] benchmark harness reliability and mid-trajectory safety, which is the kind of evaluation a golden harness (§13.2) would run; **SWE-Critix** [31] trains trajectory verifiers that could replace our rule-based proof classifier (§14, item 3). A project-level **governance manifest** for agent-mediated OSS contributions [29] and a systematisation of execution-security work for coding agents [32] frame the governance side. Table 2 summarises the differences.

**Table 2.** Closest concurrent work and how Soei differs.

| Work | Shared idea | Where Soei differs |
|---|---|---|
| Proof-or-Stop [18]; Runtime contract [19] | Agent lifecycle states are claims; proof must be fresh and bound to source state | Observation only; measures the gap on 45 real sessions (23.4% unverified, 57.8% unproved to-dos); enforcement deferred to a separate policy plane |
| Claim-aware observability [20] | Artifacts, claims and their relations as the unit of observation | Software-engineering sessions; acknowledgement tracking; content-lineage hand-offs; hash-chained history |
| OpenRath [21]; AgentTrace [23] | One unified runtime record instead of fragmented state; replay | Passive adapters over unmodified vendor agents; no framework adoption required; privacy gate before persistence |
| Skill Runtime Intelligence [24] | Passive reconstruction, graded evidence, unsupported = unknown | Scope is project-level tasks and components; Skill stages are out of scope. Adds hand-off lineage and integrity/replay properties |
| Auditable Agents [22] | Accountability requires auditability | Concrete substrate (ledger, replay, PROV graph) and measured cost (20% of raw size, 126k events/s) |
| AgentLens [25] | Passing is not the same as verified | Workplace corpus, not benchmark; episode- and to-do-level claim–proof matching with audited precision |

## 3 Gap analysis and taxonomy

We compare current project-level observability tooling (transcripts, trace dashboards and local plan/file watchers) against three requirements for a supervisory observability substrate. The gaps below are the ones we found while reviewing such tools and then confirmed in our own sessions. **(R1) Faithfulness:** what is shown is true to the evidence and its uncertainty. **(R2) Durability and reproducibility:** the same evidence yields the same view at any later time. **(R3) Evaluability:** accuracy and usefulness can be measured. The last column links each gap to the real-data measurement in §10 that quantifies it.

**Table 3.** Gap taxonomy G1–G14 with Soei requirement and empirical evidence.

| ID | Dimension | Gap and evidence | Sev. | Soei requirement | Real-data evidence (§10) |
|---|---|---|---|---|---|
| G1 | Architecture | Each view has its own service, adapters and history; no shared backend | **High** | One pipeline and ledger; views are projections | Replay determinism holds (§10.7) |
| G2 | Architecture | Views that coexist overwrite each other's state (to-do overwrite, hook collision) | **High** | Evidence-class precedence; namespaced hooks | — |
| G3 | Epistemic | Evidence classes are labels, with no confidence or fusion | **High** | Formal lattice; calibrated confidence | Audit precision 66.7→83.3% (§10.1) |

| ID | Dimension | Gap and evidence | Sev. | Soei requirement | Real-data evidence (§10) |
|---|---|---|---|---|---|
| G4 | Epistemic | Plans accepted at tool-call start | High | Two-phase acknowledgement | 1.1% phantom mutations (§10.5) |
| G5 | Epistemic | Completion = reported complete | High | Claim–proof verification | 23.4% unverified; 57.8% unproved to-dos |
| G6 | Epistemic | Hand-offs: 10 min timing or explicit receipt | High | Graded H0–H3 with content lineage | Heuristic recall 25% (§10.3) |
| G7 | Epistemic | Role inference from directory names; accuracy unmeasured | Med. | Scored attribution with abstention | Agenda (§12) |
| G8 | Temporal | In-memory/bounded history (2 MiB, 24 h) | High | Checkpointed ledger, lossless resume | 48.1% bytes beyond bound (§10.4) |
| G9 | Temporal | No cross-source causal order | Med. | HLC + source offsets | — |
| G10 | Security/privacy | Unauthenticated mutation routes on localhost; inconsistent field policies between views | High | Token, Origin/Host checks, ingest-time policy | 1,326 secret-bearing records (§10.6) |
| G11 | Platform | Windows/Cursor unvalidated; format drift | Med. | Adapter SDK + conformance corpus | Prototype ran on Windows |
| G12 | Interop | No standard export | Med. | OTel GenAI + PROV-JSON export | — |
| G13 | Supervision | Attention limited to visual cues | Low | Ranked read-only attention events | 195 returns after >30 min (§10.8) |
| G14 | Scientific | No quantitative evaluation | High | Evaluation-first harness | This paper's pilot |

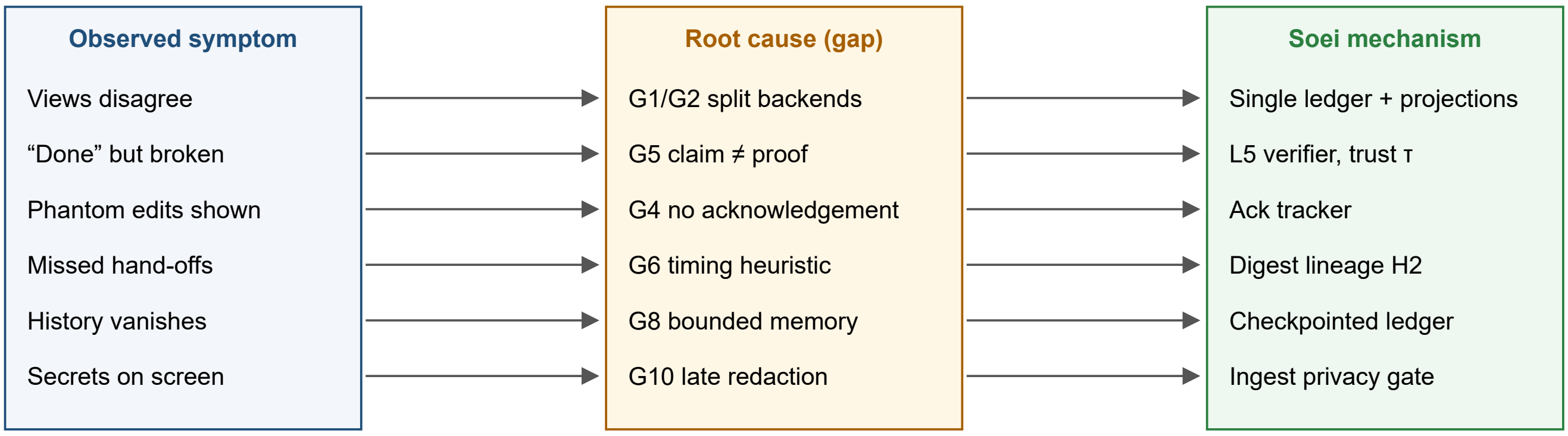

**Figure 1.** Symptom → gap → mechanism map for the high-severity gaps.

## 4 Design principles

1. **P1 Observe, never act.** Soei never edits or approves agent state, and it never prompts the agent.
2. **P2 One record, many views.** Every view is a deterministic projection of one ledger.
3. **P3 Evidence before inference.** Provenance-tagged evidence is persisted before interpretation, so inference can be re-run later with better rules and re-scored against the same evidence.
4. **P4 Claims are not proofs.** Agent-reported state is a claim. Completion is upgraded only by independent evidence.
5. **P5 Calibrated uncertainty.** Inferred relations carry measured confidence, and the system abstains when that confidence is below threshold.
6. **P6 Private by construction.** One field policy is enforced at ingest, before persistence. Secrets never enter the ledger.
7. **P7 Vendor-neutral adapters.** Provider formats are untrusted, versioned inputs validated against a conformance corpus.
8. **P8 Evaluation-first.** Recording, annotation, replay and scoring are first-class features.

## 5 Soei core architecture

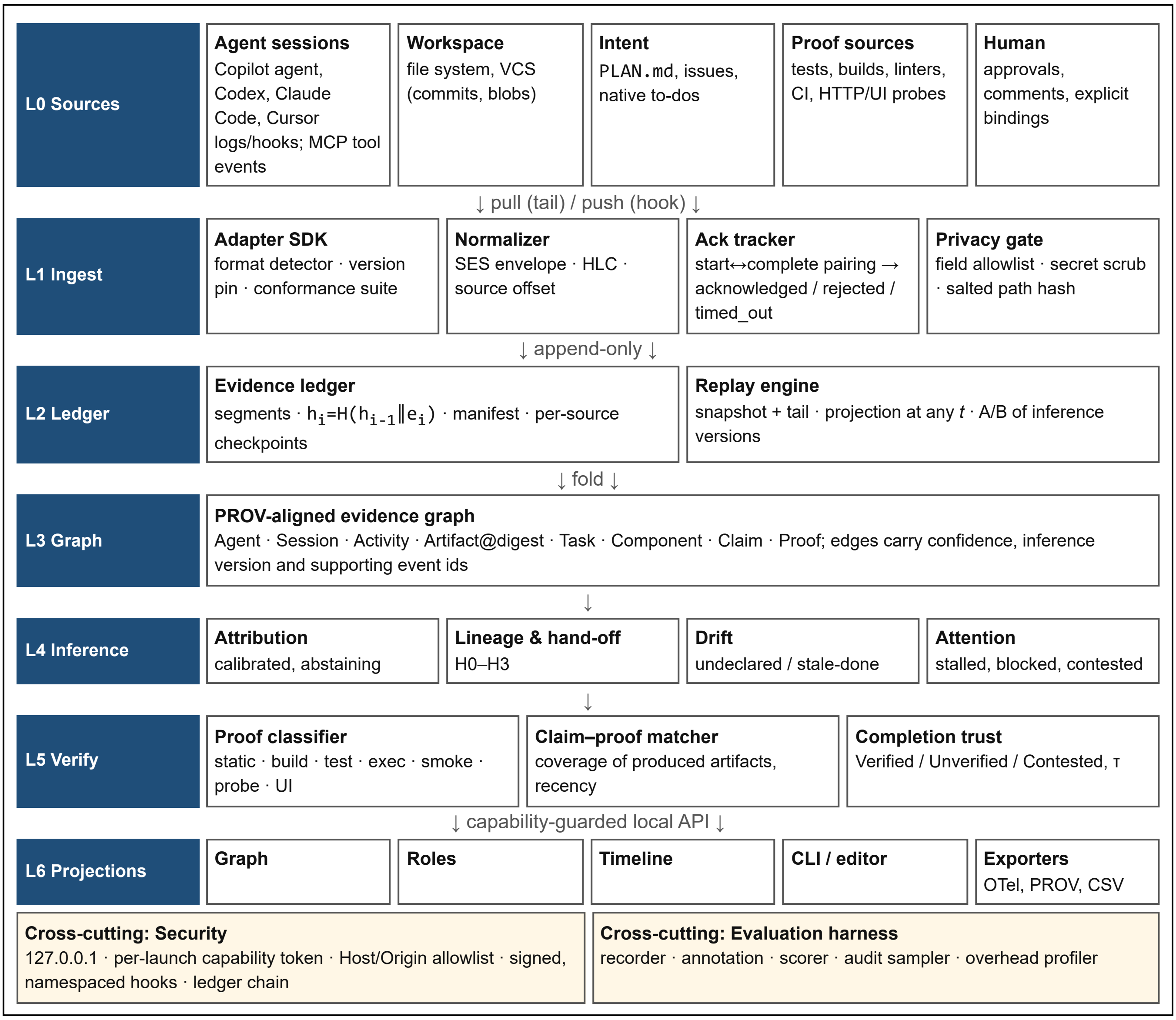

**Figure 2.** Soei layered architecture. One pipeline (L1–L5) feeds every projection (L6). Evidence is persisted before inference (P3), and verification is separate from inference (P4).

### 5.1 Core data path

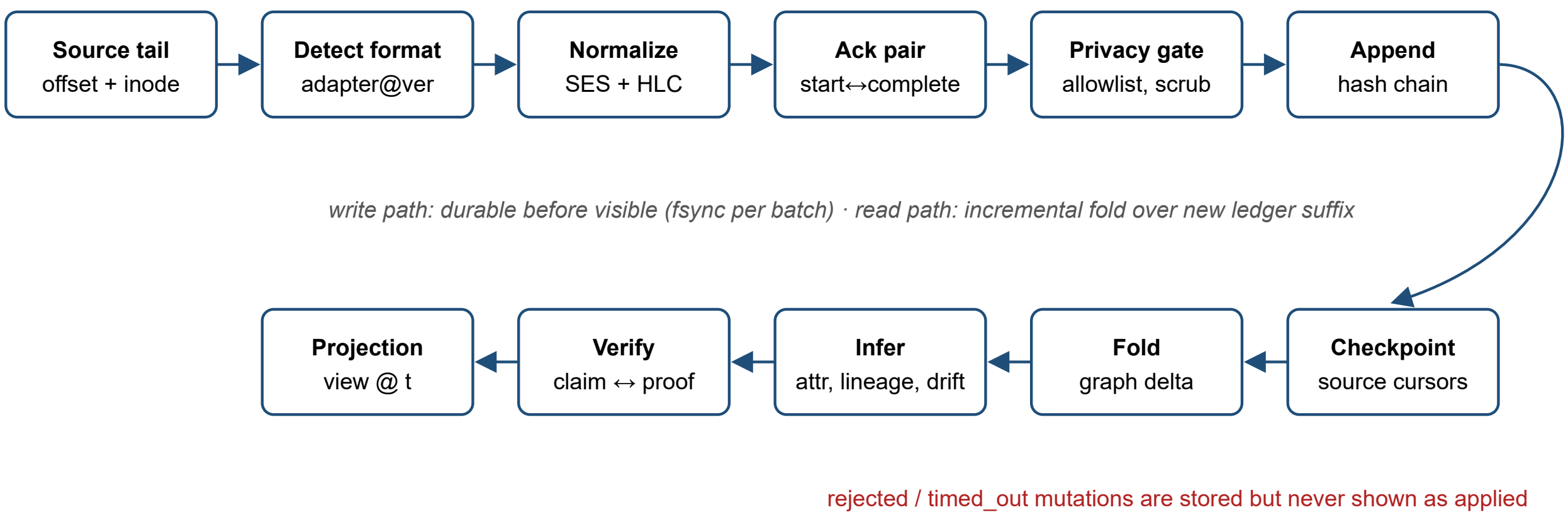

**Figure 3.** Core data path. The top row is the write path; the bottom row is the incremental read path. Evidence becomes visible only after it is durable, so the view after a crash equals the view before it.

### 5.2 L1 Ingest: adapters, Soei Event Schema and acknowledgement

**Adapter contract.** An adapter implements `detect(sample) → (format, version, confidence)`, `tail(cursor) → (records, cursor')` and `normalize(record) → SES[]`. Cursors are `(path, inode/file-id, size, byte offset)`. A rotated or truncated source is detected from the identity and size and re-anchored, never re-read blindly. Each adapter ships with a **conformance corpus** of redacted real records per supported format version. CI runs every adapter against every corpus on Windows, macOS and Linux. Unknown formats produce `evidence_class=unknown` events and are never guessed. Our prototype implements the VS Code Copilot agent adapter (session, prompt, message, tool start/complete with call ids and success flags).

**SES envelope.** CloudEvents-style [10] fields (`id`, `source`, `type`, `time`) plus `source_offset`, `hlc`, `evidence_class`, `ack`, `subject` (session, task, artifact digest, activity) and `policy` (App. A).

**Acknowledgement tracker.** Every state-changing request (edit, plan update, to-do change) is recorded as *proposed* at start and resolved by its matching completion: *acknowledged* (success), *rejected* (failure) or *timed_out* (no completion within a window or before session end). Projections render only acknowledged state by default. In our corpus this distinction affects 1.1% of mutations (§10.5).

**Privacy gate.** Runs before persistence. (i) A field allowlist per event type keeps message, prompt, reasoning, command body and tool output out by default. (ii) Remaining strings are scanned for credential patterns and replaced with typed placeholders. (iii) Paths are salted-hashed for export, with a local reversible map kept outside the ledger. Because the gate runs before the ledger, a secret can never be "un-shown" later: it is never stored.

### 5.3 L2 Evidence ledger

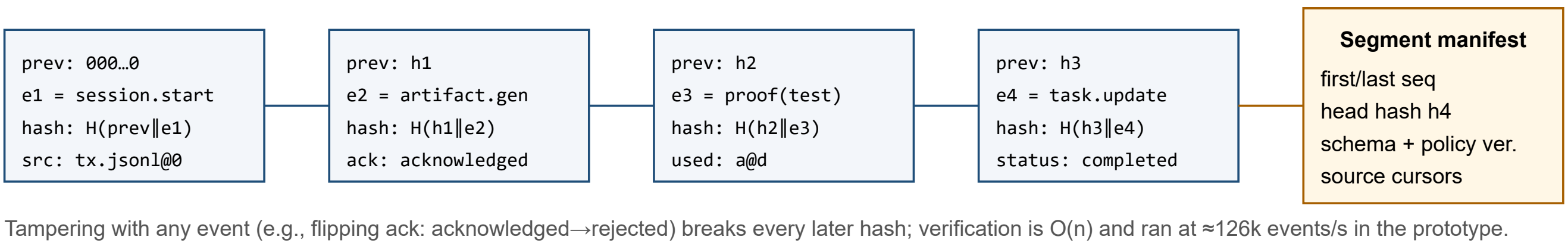

**Figure 4.** Ledger record chaining and segment manifest.

The ledger is an append-only sequence of canonical-JSON SES events grouped into segments (the prototype uses JSONL; production can use SQLite WAL with the same chain). Each record stores `prev` and `hash = SHA-256(prev ‖ canonical(e))`. A segment manifest commits to its head hash, schema and policy versions and source cursors, which gives lossless resume and integrity for shared datasets. The ledger stores **no content**: artifacts are referenced by salted path hash and, in production, content digests (BLAKE3/SHA-256 or VCS blob ids) captured at generation and use.

### 5.4 L3 Evidence graph

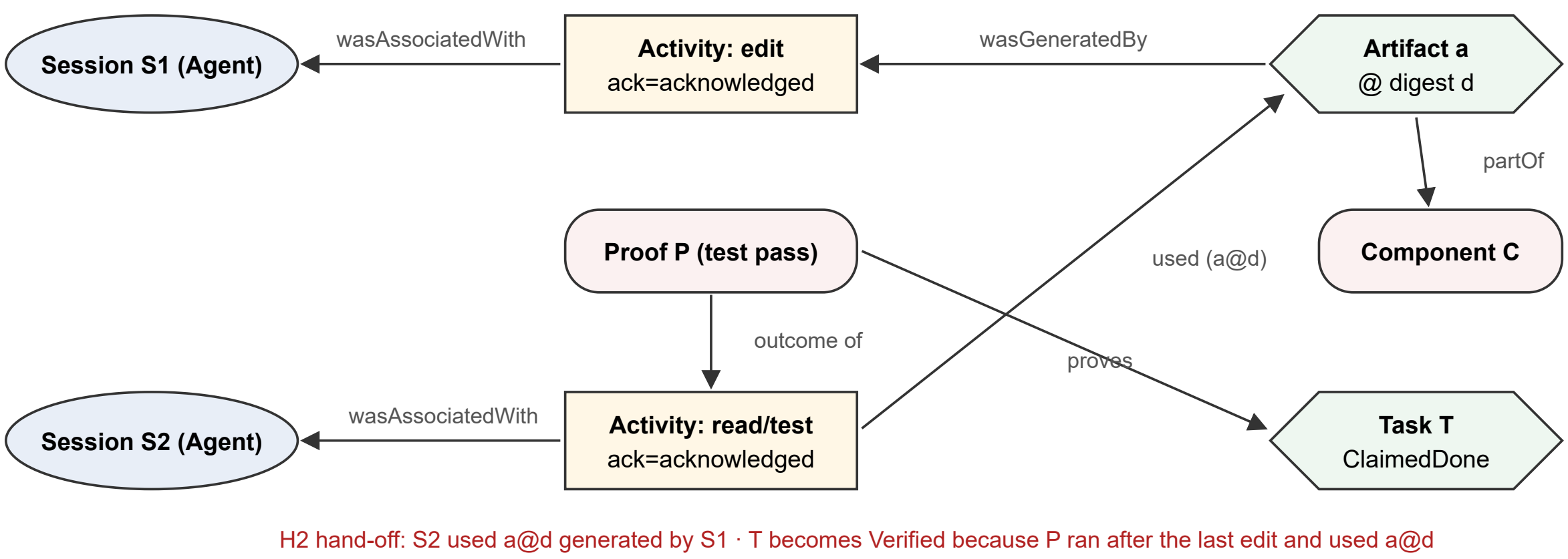

**Figure 5.** Evidence-graph schema (PROV-aligned) with an H2 hand-off and a verified claim.

The graph is folded incrementally from the ledger. Node types are *Agent, Session, Activity, Artifact@digest, Task, Component, Claim, Proof*. Edge types are *wasGeneratedBy, used, wasAssociatedWith, partOf, claims, proves, derivedFrom*. Every inferred edge carries `(confidence, inference_version, support=[event ids])`, so any rendered statement can answer "why am I seeing this?" with raw evidence.

### 5.5 L4 Inference engines

- **Attribution:** maps activities and artifacts to components and roles using plan globs, path features, co-change history and operation type. It is a calibrated classifier that abstains below threshold θ. Directory-name role heuristics become one feature among several.
- **Lineage and hand-off:** computes H0–H3 (§6.4) from *used*/*wasGeneratedBy* over artifact digests. Our data shows content lineage is required, because real reuse is slow (§10.3).
- **Drift:** undeclared and stale-done drift (§6.5) turn the usual "plan is behind" indicator into measurable signals.
- **Attention:** ranks read-only alerts: blocked-on-human, stalled task, contested claim (failing proof after a completion claim), sustained undeclared work, and returns after long idle periods.

### 5.6 L5 Verification

The proof classifier labels activities as *static* (compile, lint, type-check, diagnostics), *build* (build/package), *test* (test runners or test-named scripts), *exec* (running the changed artifact or its sibling), *smoke* (import checks), *probe* (HTTP to localhost after restart) or *UI* (browser checks after an edit). The claim–proof matcher then decides **Verified**, **Unverified** or **Contested** for every claim, and the completion trust τ summarises coverage and recency (§6.6). This is the layer the real data shows to be missing in practice (§10.1–10.2).

### 5.7 Task state machine

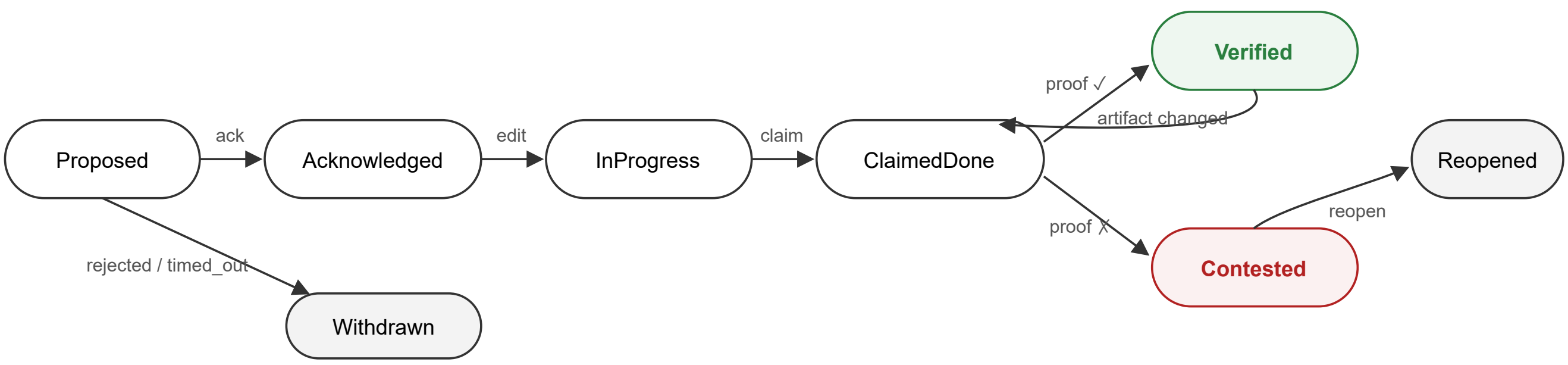

**Figure 6.** Task lifecycle. A task cannot leave *Proposed* without acknowledgement, and it cannot reach *Verified* without a proof that ran after the last edit and used the produced artifacts. Any later change returns it to *ClaimedDone*.

### 5.8 Security, trust boundaries and deployment

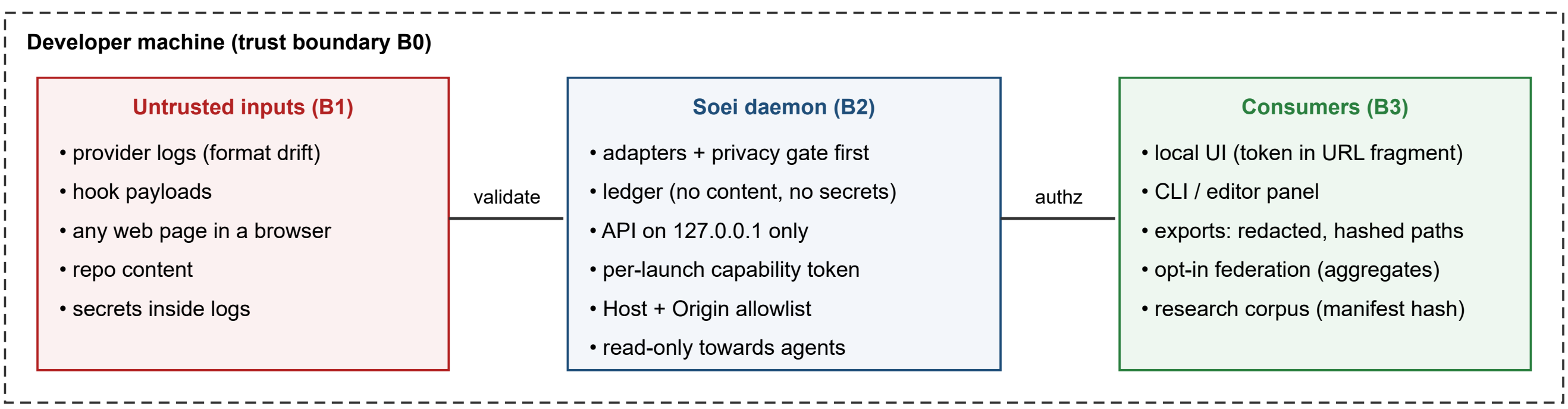

**Figure 7.** Trust boundaries. Everything entering B2 is untrusted. Persistence happens only after the privacy gate. Every API call needs the capability token and passes Host/Origin checks, which rules out the unauthenticated-localhost class of issue common in local developer tools.

Threats addressed: (T1) cross-site requests or DNS rebinding against localhost services; (T2) leakage of secrets or code through screenshots and exports; (T3) hook relay hijack or duplication; (T4) dataset tampering; (T5) adapter confusion by malformed input. Controls: 127.0.0.1 binding; a random capability token per launch, required on every route including reads; strict `Host` and `Origin` allowlists; an ingest-time allowlist and secret scrubber; signed, namespaced, idempotent hook registrations; hash-chained segments with manifests; and size and recursion limits on adapters, run under fuzzing in CI.

## 6 Formal model

> **Definition 1 (Evidence event and ledger).** An evidence event is a tuple $e = (id, src, off, \kappa, \lambda, \sigma, \alpha, \pi)$: source, source offset, kind, hybrid logical timestamp, subject references, acknowledgement state $\alpha \in \{proposed, acknowledged, rejected, timed_out, n/a\}$, and applied policy. The ledger is $L = \langle e_1 \ldots e_n \rangle$ with $h_i = H(h_{i-1} \,\|\, enc(e_i))$.

> **Definition 2 (Evidence lattice).** Statements carry a class from *Unknown* < *Inferred* < *Observed* < *Reported* < *Verified*. For statements about *intent*, Reported outranks Observed. For statements about the *world*, Observed outranks Reported. Verified is reachable only through a Proof node.

**Confidence fusion.** For a relation $r$ with independent signals of calibrated confidence $c_j$ and reliability weights $w_j$:

$$c(r) = 1 - \prod_j (1 - w_j c_j) \tag{1}$$

Competing hypotheses are normalised, and the system abstains when $max\ c < \theta$. Calibration is reported as expected calibration error.

> **Definition 3 (Task lifecycle).** *Proposed* → *Acknowledged* → *InProgress* → *ClaimedDone* → *{Verified, Contested}*, with Withdrawn and Reopened. *ClaimedDone* → *Verified* requires a proof $p$ with outcome pass, $\lambda(p) > \lambda(\text{last generation under } T)$ and $used(p) \supseteq artifacts(T)$ (or their component).

### 6.4 Hand-off levels

**Table 4.** Graded hand-off evidence between sessions $S_1$ and $S_2$.

| Level | Condition | Meaning | Measured here | |
|---|---|---|---|---|
| H0 | $S_2$ starts within Δ of $S_1$ ending | Temporal adjacency | Δ = 10 min | 7 pairs; 6 lineage-confirmed |
| H1 | H0 ∧ $S_2$ uses artifacts $S_1$ modified | Shared working set | (subsumed by H2 path-level) | |
| H2 | ∃a: wasGeneratedBy(a, $S_1$) ∧ used($S_2$, a) | Content lineage | 24 pairs, 292 links | |
| H3 | H2 ∧ explicit receipt | Confirmed transfer | N/A (no receipts in corpus) | |

### 6.5 Plan–reality drift

$$Drift_{undeclared}(W) = |O_W \setminus D_W| \,/\, |O_W| \,, \quad Drift_{stale\text{-}done}(W) = |\{C \in O_W : status(C){=}done\}| \,/\, |O_W| \tag{2}$$

### 6.6 Completion trust

$$\tau(T) = cov(T) \cdot e^{-\delta \cdot age(T)} \cdot (1 - f(T)), \quad cov(T) = |A_T \cap used(P_T)| \,/\, |A_T| \tag{3}$$

### 6.7 Properties (with empirical checks)

> **Property 1 (Projection consistency).** Views defined as functions of $G(L_{\leq t})$ agree on every shared statement. This rules out G1/G2-type divergence by construction.

**Property 2 (Replay determinism).** Given $L_{\leq t}$, inference version $v$ and policy $\pi$, the projection is deterministic. *Empirical check:* two independent replays of the real 42,398-event ledger produced identical result digests (§10.7).

**Property 3 (Tamper evidence).** Changing any event invalidates every later hash. *Empirical check:* flipping one acknowledgement in the middle of the real ledger was detected (§10.7).

**Property 4 (No silent completion).** No task reaches Verified without a Proof node, and no rejected or timed-out mutation is rendered as applied.

## 7 Algorithms

```
Algorithm 1  Acknowledgement-aware ingest
input: source records R (ordered), window W
pending ← {}
for r in R:
  if r.kind = tool.start:      pending[r.call] ← event(r, ack = proposed); emit(pending[r.call])  # not rendered
yet
  elif r.kind = tool.complete: e ← pending.pop(r.call); e.ack ← acknowledged if r.success else rejected
                               emit_resolution(e)                                        # now renderable
  else: emit(normalize(r))
on session end or age > W:   for e in pending: e.ack ← timed_out; emit_resolution(e)
```

```
Algorithm 2  Claim-proof matching (per episode and per task)
for each episode / task T:
  lastEdit ← max λ over acknowledged artifact.generated under T
  proofs   ← { p : p.kind ∈ {static, build, test, exec, smoke, probe, ui}, p.ack = acknowledged, λ(p) > lastEdit
}
  if proofs = ∅                      : state(T) ← Unverified
  elif ∃p ∈ proofs failed covering A_T  : state(T) ← Contested
  else                                 : state(T) ← Verified; τ(T) ← Eq.(3)
```

```
Algorithm 3  Lineage hand-off detection (H2)
writes[a] ← list of (λ, session) for acknowledged generations of artifact a   # a = path hash (prototype) or
path@digest
for each acknowledged use (a, λu, S2):
  (λw, S1) ← latest write of a with λw ≤ λu
  if S1 ≠ S2: link(S1 → S2, gap = λu − λw)               # H2 if digest(write) = digest(use)
pairs ← distinct (S1, S2) over links; compare with H0 pairs (end→start ≤ 10 min)
```

**Complexity.** Ingest and chain verification are O(n) in events. Lineage is O(u log w) with per-artifact sorted write lists. Claim–proof matching is O(n) with one pass per session. All prototype stages ran in seconds on the full real corpus (§10.7).

## 8 Implementation blueprint

**Table 5.** Reference implementation choices.

| Concern | Choice | Prototype status |
|---|---|---|
| Runtime | Single local daemon (Python 3.11+ or Node 20+) | Python, stdlib only |
| Ledger | Canonical JSON + SHA-256 chain; JSONL or SQLite WAL segments | JSONL, implemented |
| Acknowledgement | Start/complete pairing by call id | Implemented |
| Privacy gate | Allowlist + secret scrub + salted path hashes | Allowlist + hashing implemented; scrub validated (0 matches in ledger) |
| Proof classifier | Versioned rules + learned model later | Rules v1–v4, audited |
| Lineage | Content digests at generation/use; Git blob ids | Path-hash level (digest is future work) |

| Concern | Choice | Prototype status |
|---|---|---|
| Clocks | Hybrid logical clocks per source | Wall time + source offsets |
| API | HTTP + SSE, 127.0.0.1, capability token, Host/Origin allowlist | Specified |
| Interop | OTel GenAI + PROV-JSON exporters | Specified |

```
soei/
  adapters/   copilot_vscode/ codex/ claude/ cursor/ fs/ git/ junit/ ci/   (+ conformance/<format>/<version>/)
  core/       ses.schema.json  normalizer  ack_tracker  hlc  privacy_gate
  ledger/     writer  segments  manifest  checkpoints  verify_chain  replay
  graph/      prov_model  fold  snapshot  query
  inference/  attribution  lineage  drift  attention  calibrate
  verify/     proof_classifier  claim_proof  trust
  api/        server(token, origin)  sse  exporters/{otel, prov, csv}
  views/      graph  roles  timeline  cli  vscode
  eval/       recorder  audit_sampler  annotate  score  overhead  ab_replay
```

**Table 6.** Roadmap.

| Phase | Deliverable | Gaps | Exit criterion |
|---|---|---|---|
| 1 Substrate | SES, ledger, checkpoints, Copilot/Codex/Claude/FS/Git adapters, ack tracker, secure API | G1 G2 G4 G8 G9 G10 | Lossless restart; chain verifies; zero view divergence |
| 2 Graph & inference | PROV graph, digest lineage, calibrated attribution, drift | G3 G6 G7 | Macro-F1 and ECE on an annotated corpus |
| 3 Verification | Proof classifier v5 (learned), test/CI adapters, τ | G5 | Audit precision ≥ 90% both labels |
| 4 Reach | Cursor/Windows conformance, exporters, VS Code panel, attention | G11 G12 G13 | Conformance passes on 3 OSes |
| 5 Study | Multi-developer corpus, user study, release | G14 | §12 protocol executed |

## 9 Empirical pilot study (real data)

### 9.1 Research questions

1. **PQ1 (Claim–proof):** How often does an agent finish an episode, or mark a to-do complete, after editing code but without independent verification?
2. **PQ2 (Hand-offs):** How much real cross-session artifact reuse does a timing heuristic capture compared with content lineage?
3. **PQ3 (History bounds):** How much real evidence falls outside bounded discovery and replay windows?
4. **PQ4 (Acknowledgement):** How often are state-changing requests not acknowledged?
5. **PQ5 (Privacy):** How much sensitive material do raw agent logs contain, and does an ingest-time gate keep it out?
6. **PQ6 (Cost and integrity):** What does a hash-chained ledger cost, and do determinism and tamper evidence hold on real data?

### 9.2 Corpus

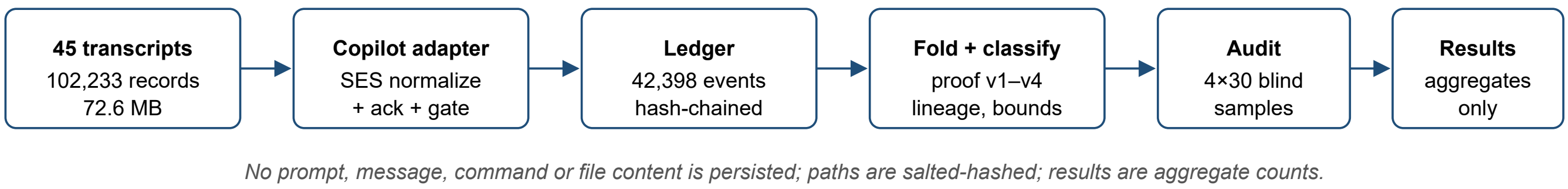

**Figure 8.** Pilot pipeline: the Soei prototype applied to the real corpus.

The corpus is every VS Code Copilot agent-mode transcript available on the first author's workstation: **45 sessions**, 23 April to 5 October 2026, **102,233 raw records** (72.6 MB), **20,577 tool executions** over 45+ tool types, **2,336 human-to-agent episodes** (an episode runs from one human prompt to the next), of which **1,044** contained file edits and **893** edited code. Work spanned post-silicon validation content (Python/PowerShell), web portals, VS Code extensions, data pipelines and documentation. All sessions were real work, not tasks constructed for the study.

### 9.3 Measures

- **Unverified code episode:** an episode with ≥1 acknowledged code edit and no acknowledged proof activity (§5.6) after its last code edit.
- **Unproved completion:** a to-do that turned *completed* after acknowledged edits made while it was *in-progress*, with no proof after those edits and before completion.
- **Lineage link:** session $S_2$ read artifact $a$ whose latest prior acknowledged write was by $S_1 \neq S_2$ (path level). **H0 pair:** $S_2$ started within 10 min of $S_1$ ending.
- **Phantom mutation:** an edit request whose completion was a failure or missing.
- **Secret exposure:** raw records matching credential patterns (GitHub, AWS and Slack token formats, private-key headers, password assignments). Counted only; values are never stored or printed.

### 9.4 Classifier audit protocol

The proof classifier is a set of versioned rules. To avoid reporting an artefact of the heuristics, the first author ran **blind audits**. For each version a seeded random sample of 30 episodes carrying the label was drawn, and the actions after the last code edit (tool names and truncated commands only) were inspected to decide whether an independent check really happened. False negatives drove each revision: **v2** added build commands, quoted interpreter paths, import smoke checks and test-named scripts; **v3** excluded deploy/sync invocations; **v4** added localhost HTTP probes, browser UI checks, invocation of edited PowerShell scripts and packaging. Every version is re-run on the whole corpus as a sensitivity analysis, and the final estimate is corrected for measured precision in both directions.

Two of my own misreadings (first author) shaped these rules more than any design discussion did. In the v1 audit I had counted `python -c` import checks and scripts named `_tmp_test_*` as "no verification" because they did not look like tests; in this corpus they are one of the main ways an agent checks its own work. The opposite mistake survived to v4: one `scp` deploy was read as execution of the changed script, and only reading the raw tool sequence showed that nothing had run locally. I would not have found either by looking at aggregate rates.

## 10 Results

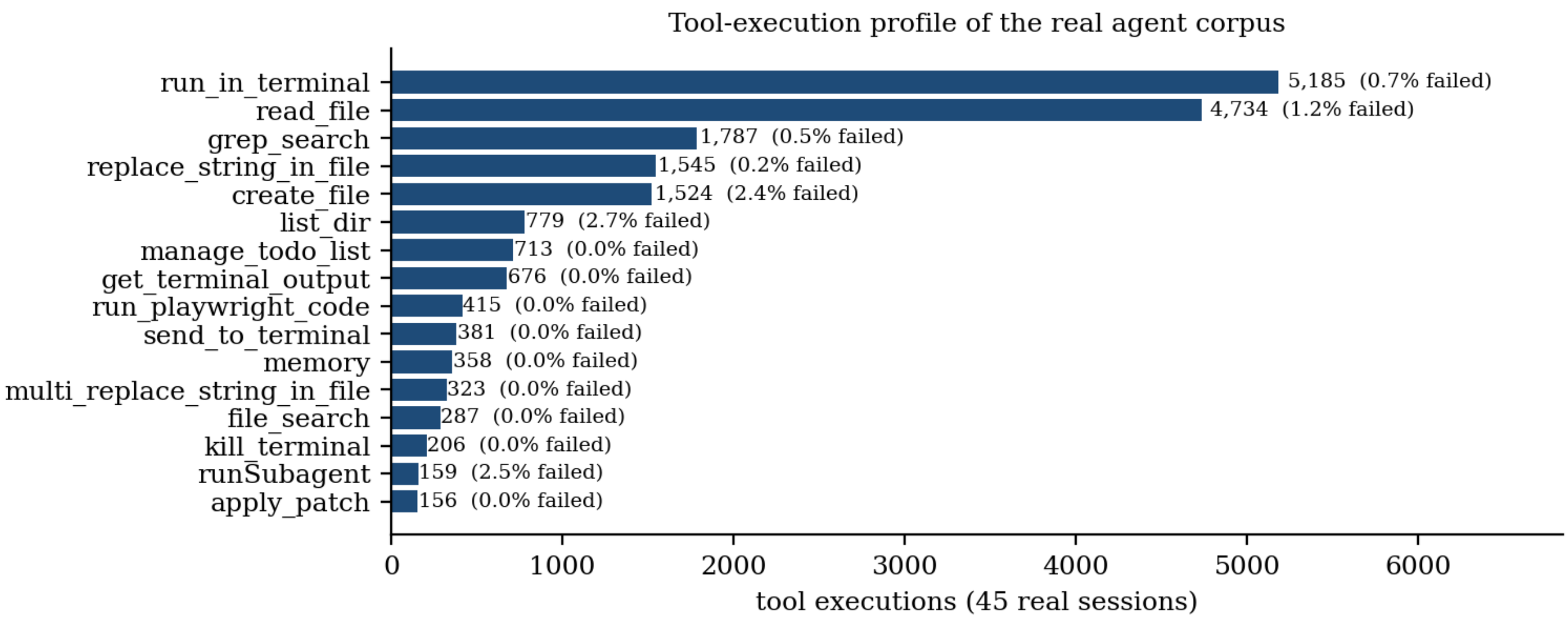

**Figure 9.** Tool-execution profile of the real corpus (top 16 tools) with failure rates. Terminal use and file reading dominate, and edits are frequent (3,548 mutation requests).

### 10.1 PQ1: the claim–proof gap

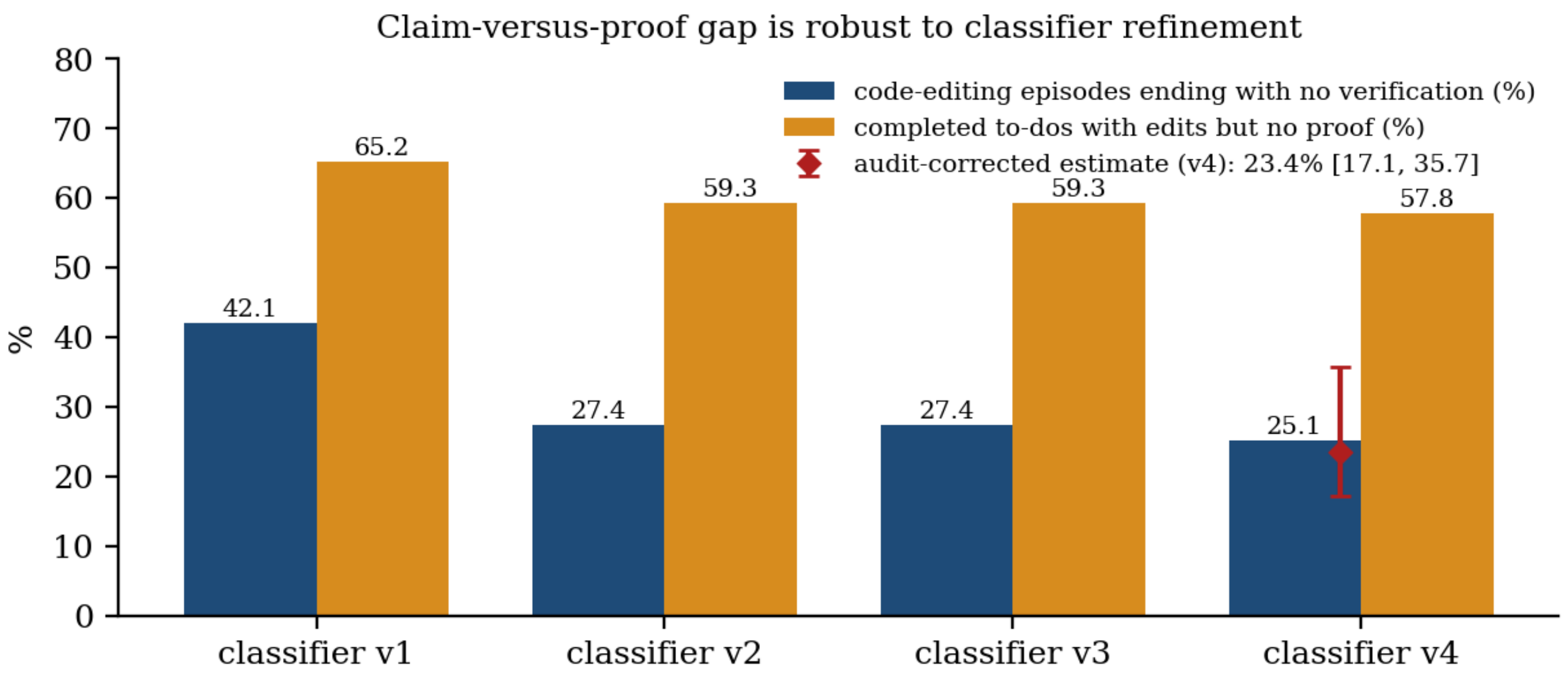

**Figure 10.** Share of code-editing episodes that end unverified (blue) and of edit-backed to-dos completed without proof (orange) across classifier versions. The red marker is the audit-corrected estimate with a 95% CI.

**Table 7.** Claim–proof results by classifier version (real corpus; 893 code-editing episodes; 322 edit-backed completed to-dos).

| Version | Unverified code episodes | % | Unproved completions | % | Audit precision ('unverified') |
|---|---|---|---|---|---|
| v1 | 376 | 42.1 | 210 | 65.2 | 20/30 = 66.7% |
| v2 | 245 | 27.4 | 191 | 59.3 | — |
| v3 | 245 | 27.4 | 191 | 59.3 | 22/30 = 73.3% |

| Version | Unverified code episodes | % | Unproved completions | % | Audit precision ('unverified') |
|---|---|---|---|---|---|
| v4 (final) | 224 | 25.1 | 186 | 57.8 | 25/30 = 83.3% |

With v4 precision for the "unverified" label $p_U = 0.833$ and for the "verified" label $p_V = 0.967$ (29/30), the corrected prevalence is

$$\hat{\pi} = \ell_U \cdot p_U + (1 - \ell_U)\cdot(1 - p_V) = 0.251\cdot0.833 + 0.749\cdot0.033 = \mathbf{23.4\%} \quad [\text{Wilson-propagated 95\% CI: 17.1\%, 35.7\%}] \tag{4}$$

**Finding 1.** In real day-to-day use, about **one in four** code-changing interactions ends with the agent handing control back without any independent check after its last edit. The rate falls as the classifier improves (42.1 → 25.1%), but it settles well above zero. The bias-corrected lower bound is 17.1%.

**Finding 2.** **57.8%** of to-dos that the agent itself marked *completed* after editing had no verification between edit and completion. A dashboard that renders "completed" to-dos as delivered work would show these as done. This is direct evidence for gap G5 and for Soei's claim–proof layer.

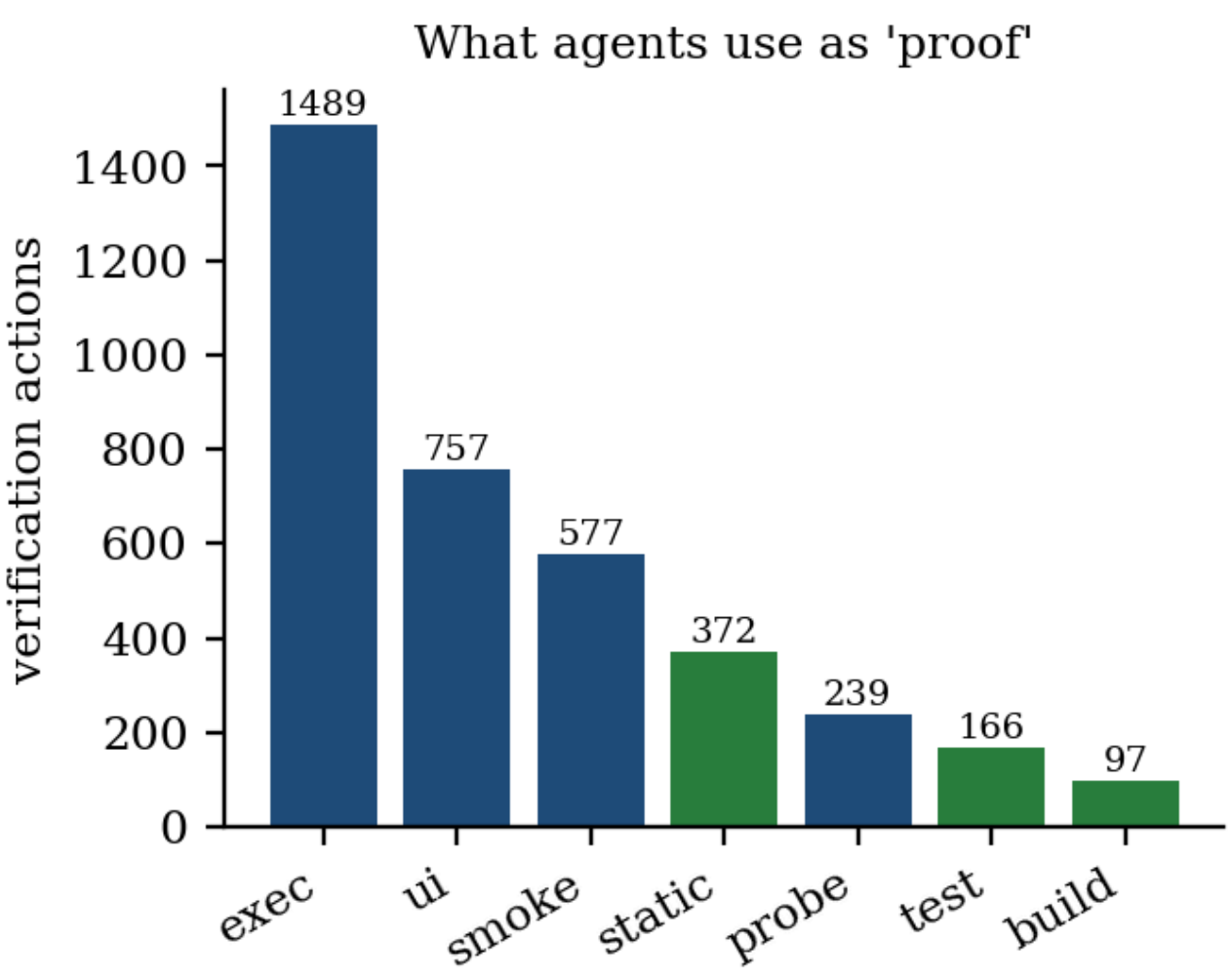

**Figure 11.** What agents use as "proof" (v4). Executing the changed artifact and UI checks dominate. Formal tests are rare (166 of 3,697 verification actions, 4.5%).

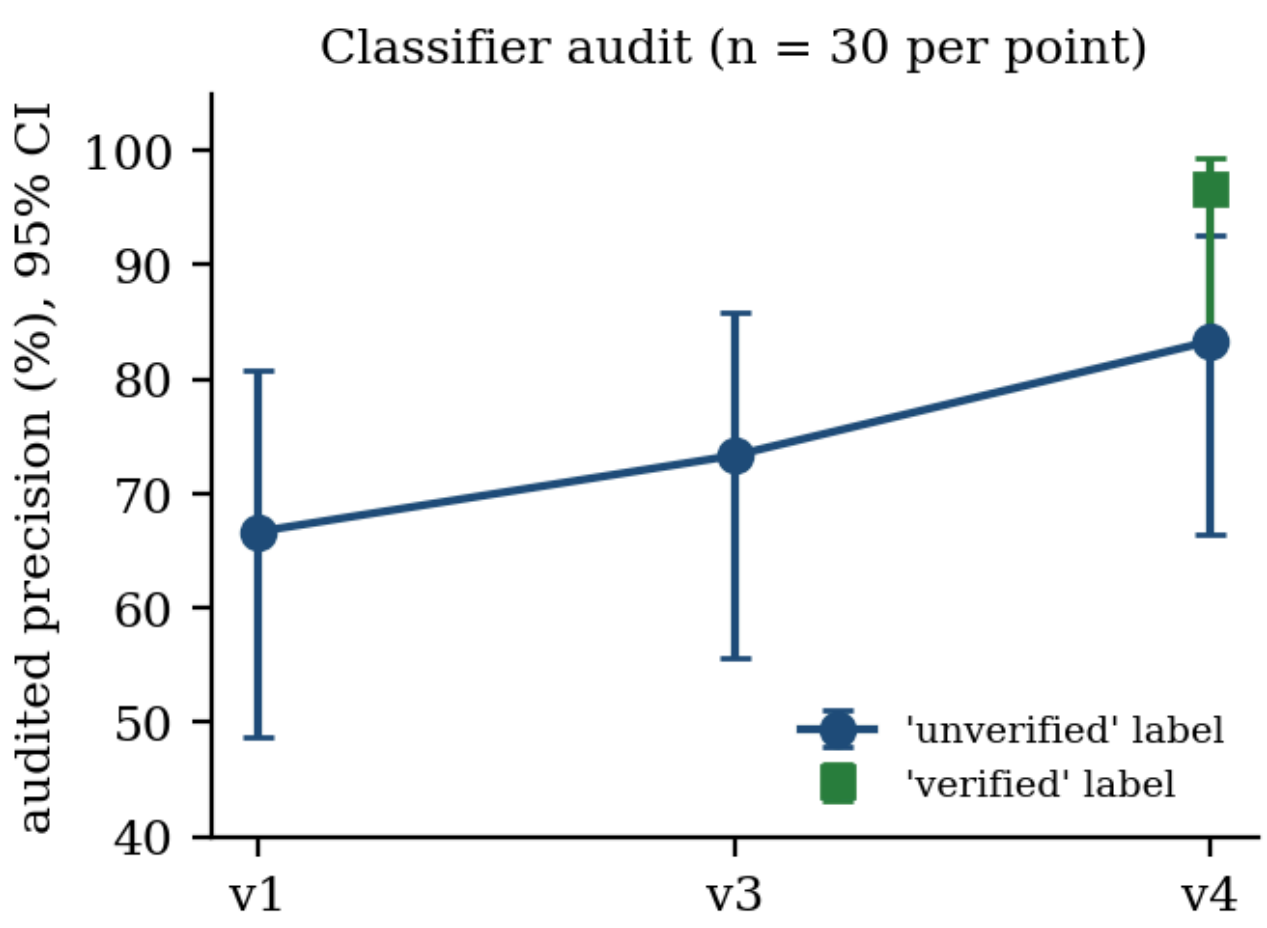

**Figure 12.** Audited classifier precision with 95% Wilson intervals, n = 30 per point.

**Finding 3.** When agents do verify, they mostly *run the changed thing* (exec 40.3%, UI 20.5%, smoke 15.6%); formal *tests* account for 4.5% of verification actions. Verification in practice is behavioural and ad hoc, so Soei's proof taxonomy must cover execution, probe and UI evidence as well as test reports.

### 10.2 Agent-reported completion versus independent evidence

Taken together, Findings 1–3 show that "reported complete" and "observed verified" are different quantities in real use, and that the difference is large enough to change a supervisor's decision. Soei renders the two separately: **Verified** (green), **Unverified** (amber) and **Contested** (red), with τ as a secondary cue.

### 10.3 PQ2: hand-offs are slow, so timing heuristics miss them

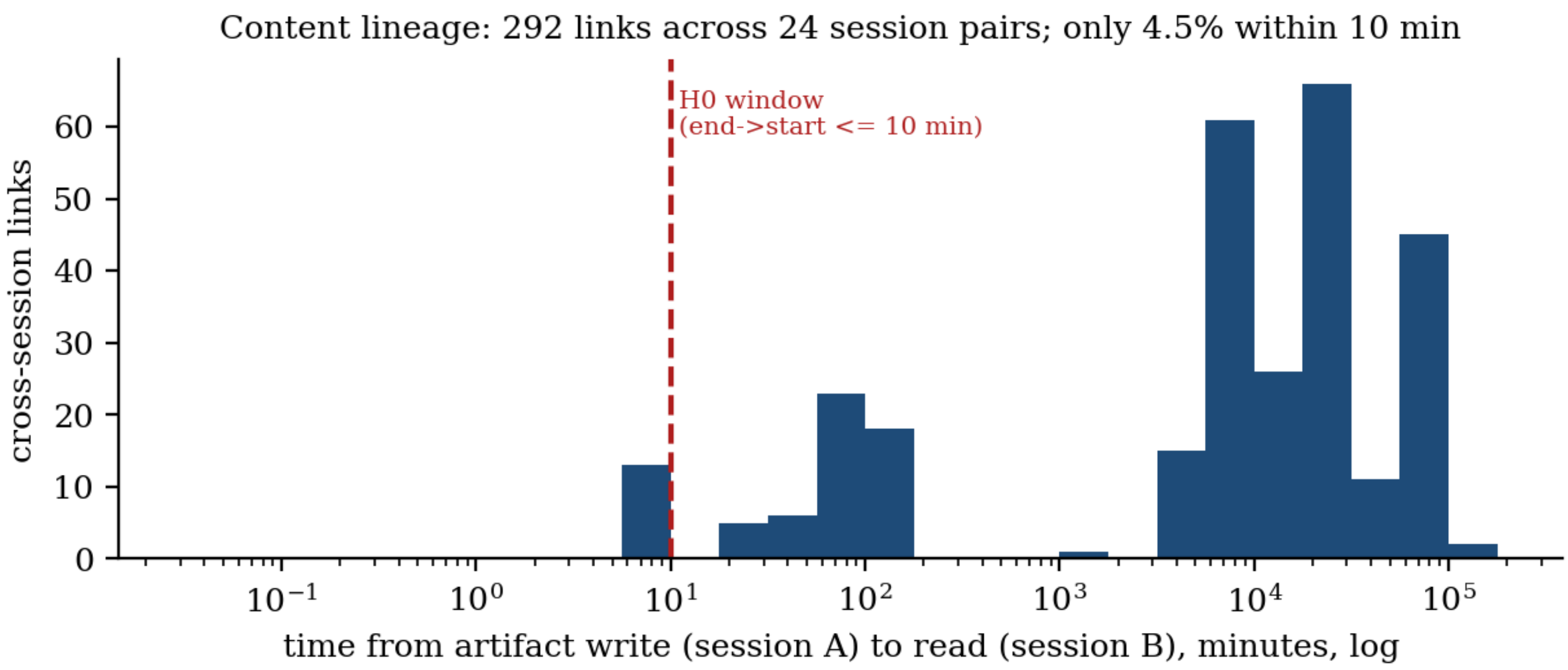

**Figure 13.** Gap between the latest write of an artifact in session A and its read in session B, over 292 real cross-session lineage links. The red line is the 10-minute window of the temporal hand-off heuristic (H0).

**Table 8.** Content lineage versus temporal heuristic (H0, Δ = 10 min).

| Quantity | Value |
|---|---|
| Cross-session lineage links / distinct session pairs | 292 / 24 |
| Links whose write→read gap is ≤ 10 min | 4.5% |
| Median write→read gap | 12,742 min (≈ 8.8 days) |
| Session pairs flagged by H0 (end→start ≤ 10 min) | 7 |
| H0 pairs confirmed by lineage (precision) | 6 / 7 = 85.7% |
| Lineage-linked pairs recovered by H0 (recall) | 6 / 24 = 25.0% |

**Finding 4.** Temporal hand-off detection is **precise but blind**: when it fires it is usually right (85.7%), but it recovers only **25%** of the session pairs that actually build on each other's artifacts. Real reuse happens across days (median 8.8 days, bimodal around hours and weeks), so only content lineage (H2) can capture multi-session collaboration.

### 10.4 PQ3: bounded history loses most of the evidence

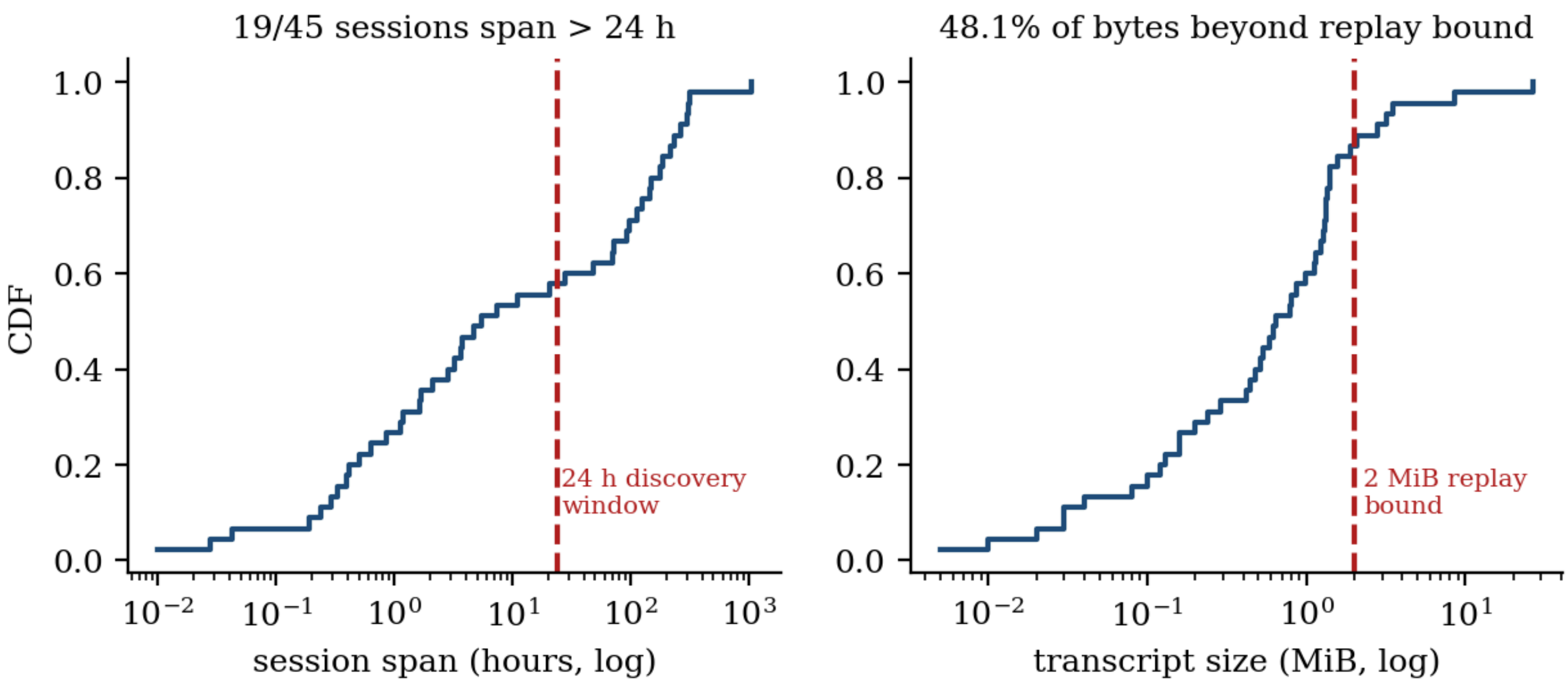

**Figure 14.** Left: CDF of session span (19/45 exceed a 24 h discovery window). Right: CDF of transcript size against a 2 MiB replay bound.

**Finding 5. 42%** of real sessions span more than 24 h (median 5.5 h, max ≈ 43 days), and **6 of 45** transcripts exceed 2 MiB. Those six hold **48.1% of all bytes**, which a 2 MiB initial replay would not reconstruct after a restart. Long, valuable sessions are exactly the ones bounded history drops, which motivates a checkpointed ledger (G8).

### 10.5 PQ4: acknowledgement

Of **3,548** edit requests, **40 (1.1%)** were not acknowledged as successful, spread across **30** episodes. Overall, 193 of 20,577 tool executions failed (0.9%). Failure rates were highest for `create_file` (2.4%), `list_dir` (2.7%) and interactive questions (6.0%).

**Finding 6.** An observer that records state changes at request time (as some legacy plan adapters do) would show about **1 in 90** edits as applied when they never happened. The rate is small, but each error is silent and propagates into every downstream inference. Two-phase acknowledgement removes it at essentially no cost.

### 10.6 PQ5: raw agent logs are a privacy liability

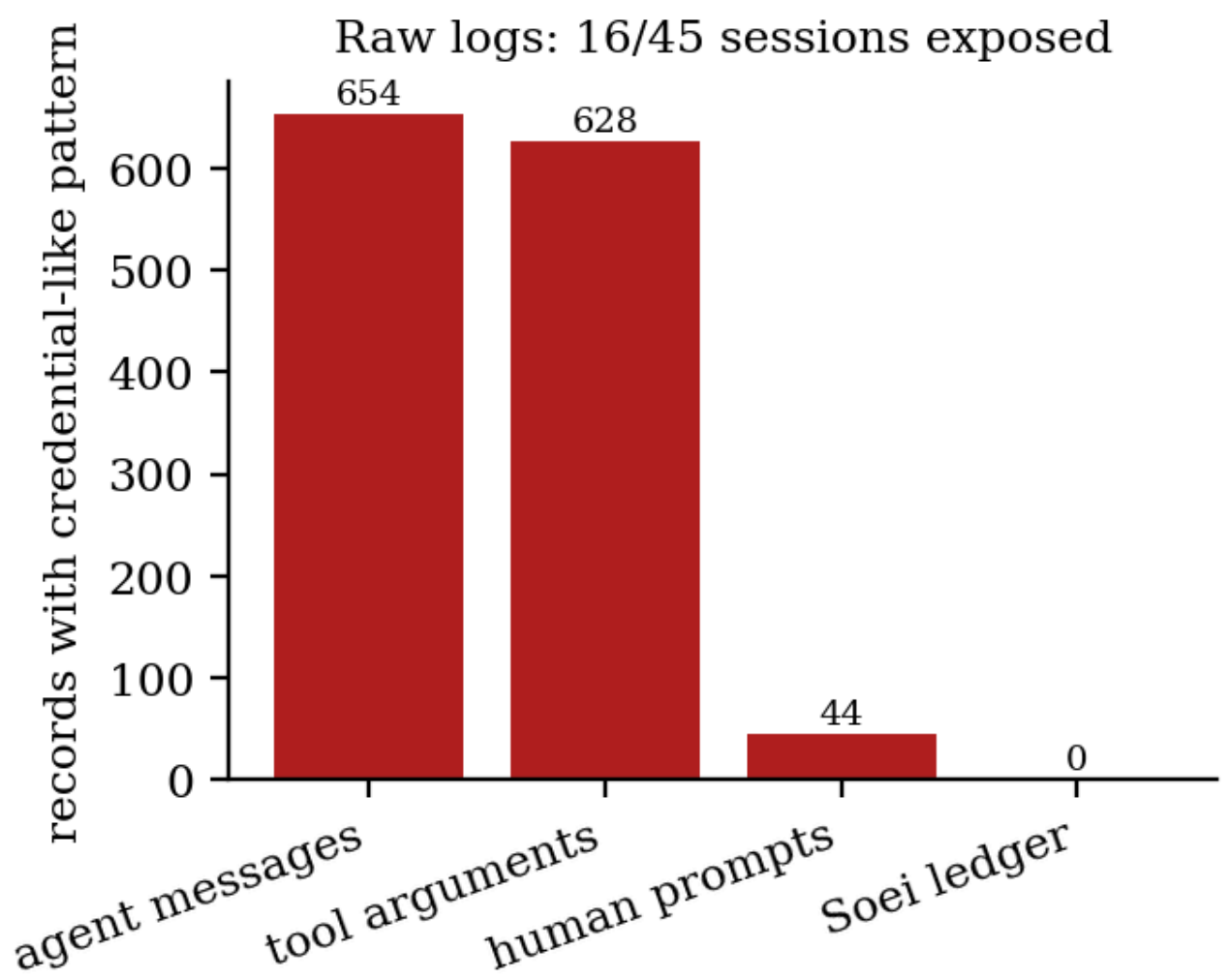

**Figure 15.** Raw records with credential-like patterns, by record type, compared with the Soei ledger.

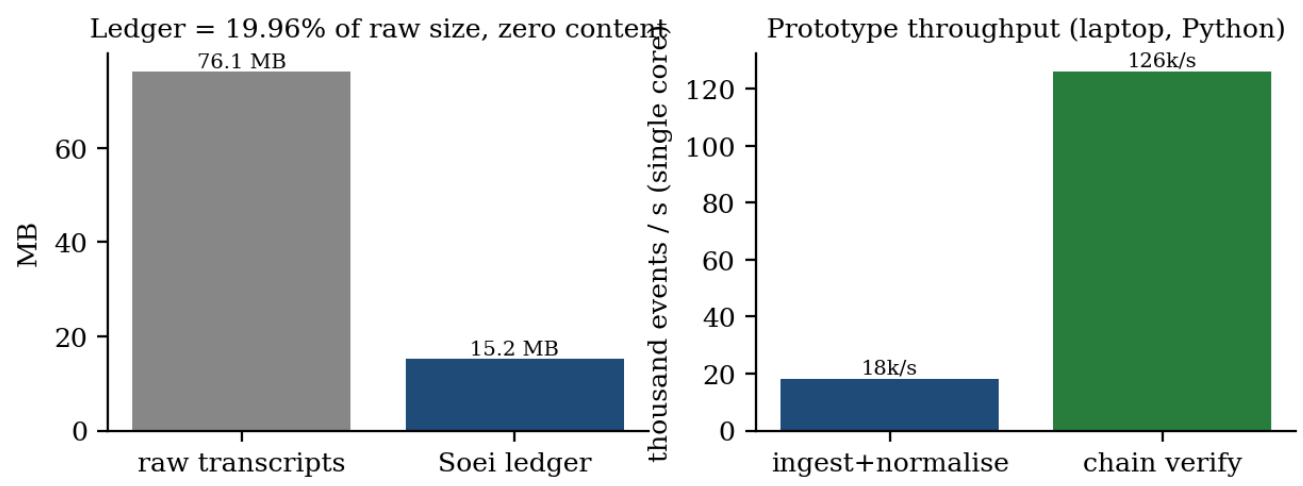

**Figure 16.** Ledger size and prototype throughput on the real corpus.

**Finding 7.** **1,326** raw records in **16 of 45** sessions contain credential-like strings, including **208 high-confidence token matches** (5 distinct tokens, 11 sessions). They appear in human prompts (44), agent messages (654) and tool arguments (628): agents repeat secrets in their own messages and pass them into commands. Tools that display tool-call details can therefore show secrets on screen and in recordings. The Soei ledger built from the same corpus contains **0** matches, because the privacy gate runs before persistence. *(The affected credentials must be rotated before any wider use of the corpus.)*

### 10.7 PQ6: cost and integrity

**Table 9.** Prototype cost and integrity on the real corpus (single laptop core, CPython).

| Measure | Value |
|---|---|
| Raw transcripts → ledger | 76.1 MB → 15.2 MB (19.96%) |
| Ledger events | 42,398 |
| Ingest + normalise + gate | 2.34 s (≈18k events/s) |
| Ledger write / full chain verification | 0.45 s / 0.34 s (≈126k events/s) |
| Graph fold + all analyses | 0.62 s |
| Replay determinism (two independent replays) | identical digests ✓ |
| Tamper test (flip one ack mid-ledger) | detected ✓ |
| Content / secret matches stored | 0 / 0 |

Provenance turned out to be cheap. Five months of one developer's agent activity fits in a 15 MB tamper-evident ledger that verifies in a third of a second and replays deterministically. At this scale the overhead argument against durable, verifiable agent history does not hold (Finding 8).

### 10.8 Supervision patterns

The corpus contains **195** human returns after more than 30 minutes of session inactivity (median gap 150 min). Each is an instance of the thirty-minute question: what happened while I was away, and is it done? Each is also a moment when Soei's attention model would present Unverified and Contested claims first.

## 11 Discussion and threats to validity

**What the data changes.** Three design decisions that were previously argued now have measurements behind them. (1) *Verification is not optional:* a quarter of code-changing episodes and over half of agent-completed edit-backed to-dos lack proof. (2) *Lineage must be content-based:* timing heuristics see a quarter of real collaboration. (3) *History must be durable and private:* long sessions carry most of the evidence and many secrets.

**Internal validity.** The proof classifier is rule-based. We mitigate this in two ways: blind audits (4 samples of 30) with a two-sided precision correction and propagated confidence intervals, and a sensitivity analysis across four classifier versions. The audit also changed our view of what counts as proof. Running the edited script, hitting localhost or opening the page in a browser is how agents check work in this corpus, and test runners are 4.5% of verification actions; a classifier built from unit-test conventions alone would have overstated the gap by almost half (42.1% at v1 against 25.1% at v4). Proofs in other channels (e.g., the human testing manually outside the agent) are invisible, which biases the "unverified" rate upwards from a *human* perspective but not from an *agent-evidence* perspective. The latter is what an observability tool can display. Lineage uses path-level identity, not content digests: overwrites between write and read could create spurious links, and digest-level lineage is planned. To-do completion uses the agent's own to-do tool, whose granularity varies.

**External validity.** The corpus comes from one developer and one agent product (VS Code Copilot agent mode, which ran several frontier models), in a post-silicon validation and tooling context. Rates will differ across teams, agents and domains. We therefore frame the results as an *existence and magnitude pilot* that motivates design, not as population estimates. AgentLens [25] reports missing or disordered verification in 10.7% of passing benchmark trajectories with a different agent and corpus, which suggests the phenomenon is not specific to our setting. The multi-developer study in §12 addresses generalisation.

**Construct validity.** "Verified" means an independent check ran after the last edit, not that the code is correct. Soei never presents $\tau$ as a correctness probability.

**Ethics and privacy.** All analysis ran locally on the authors' own sessions. No message, prompt, command or file content was persisted, and only aggregate counts are reported. Credential values were never printed or stored, and rotation of the affected credentials is required. A multi-developer study requires informed consent and ethics review.

## 12 Research agenda

**Table 10.** Follow-on studies.

| RQ | Question | Design | Metrics |
|---|---|---|---|
| RQ1 | Attribution fidelity and calibration | Annotated multi-repo corpus; two annotators, $\kappa$ | Macro-F1, coverage at $\theta$, ECE |
| RQ2 | Digest lineage versus path lineage versus H0 | Instrumented digests at write/read; seeded hand-offs | P/R/F1 per level |
| RQ3 | Does claim–proof display reduce false acceptance? | Within-subjects, ≥16 developers, replayed sessions with seeded false completions | False-acceptance rate, decision accuracy |
| RQ4 | Integrity and cost at team scale | Fault injection (kill −9, rotation, format change) | Event loss, divergence, recovery time, p95 latency |
| RQ5 | Situation awareness | SAGAT probes [4], counterbalanced: transcripts versus a trace dashboard versus Soei | Probe accuracy, time-to-awareness, NASA-TLX, SUS |
| RQ6 | Generalisation of the pilot findings | Opt-in corpus from multiple developers, agents (Codex, Claude, Cursor) and domains | Rates of Findings 1–7 with CIs per stratum |

**Baselines and ablations.** Baselines: raw transcripts, an LLM trace dashboard, a local plan/file watcher. Ablations: Soei without acknowledgement, without lineage (H0 only), without verification, without calibration, and with directory-name roles only. Replay determinism lets every ablation run on identical evidence.

## 13 Why this is required: the golden harness and governance of autonomous agents

### 13.1 The autonomy problem

Every episode in our corpus had a human at the end of it, and still about one in four code-changing episodes ended unverified (§10.1). In that setting the human is the last line of defence, re-reading and re-running what the agent left behind. As agents become *autonomous*, running for hours, chaining sub-agents, merging and deploying without per-step approval, that defence goes away. An unverified hand-back is then a defect escape path. A missed hand-off becomes an unaccountable change and a phantom edit a corrupted audit trail. A secret in a log becomes an incident. Autonomy therefore needs a substrate that can answer, for any change and at any later time, four governance questions (compare the auditability dimensions of [22]):

1. **Accountability:** which agent, session and model produced this exact artifact version, and on whose instruction?
2. **Assurance:** what independent evidence shows the change works, and is that evidence newer than the change?
3. **Auditability:** can the record be shown to be complete and untampered, and can the decision context be replayed exactly?
4. **Containment:** did any sensitive data enter the record or leave the machine?

Soei's layers map one-to-one onto these questions: lineage and attribution (L3–L4) for accountability, the claim–proof verifier (L5) for assurance, the hash-chained ledger and replay (L2) for auditability, and the ingest-time privacy gate (L1) for containment.

### 13.2 The golden harness

We use **golden harness** for a reference evaluation and assurance environment in which (i) every agent action is captured as verifiable evidence, (ii) every completion claim is checked against independent proof or a task oracle, and (iii) every run can be replayed and compared bit-for-bit across agent, model or prompt versions. It serves three purposes: *evaluation* (how good is this agent on our work?), *regression* (did the model or agent upgrade make it worse or less honest?) and *certification* (is this agent fit for autonomy level $L_k$ on this class of task?). Harness benchmarks such as TraceTrial-Bench [27] and HarnessAudit [28] supply the task and safety side; evidence-gated lifecycle control [18, 19] supplies the enforcement side. Soei supplies the record both consume.

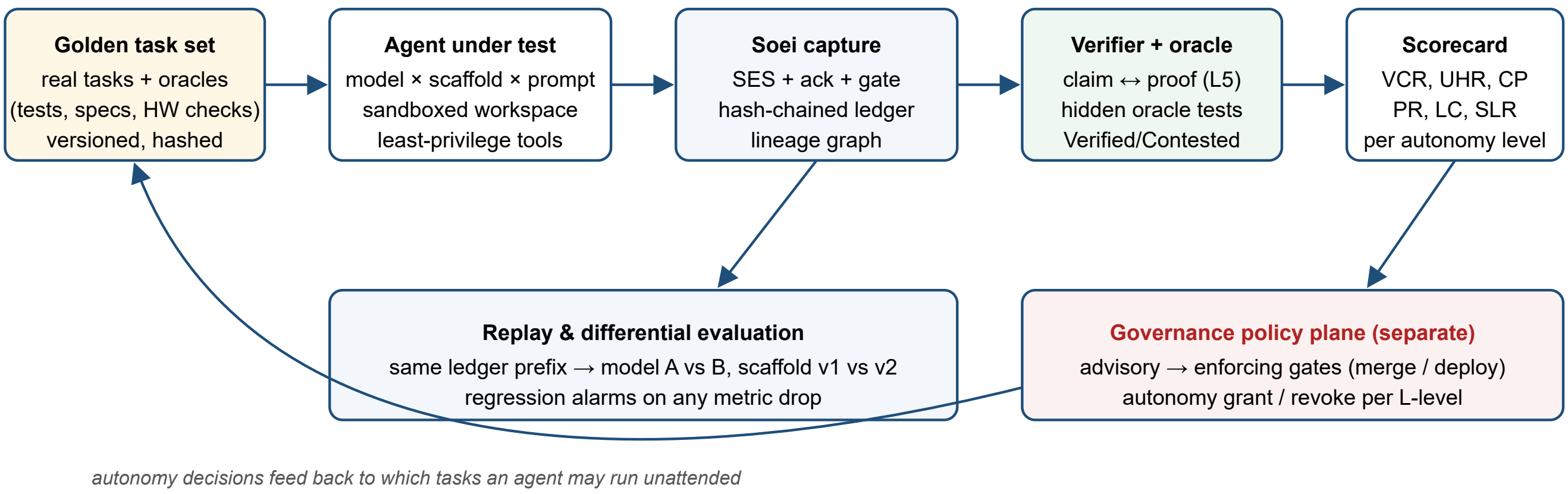

**Figure 17.** Golden harness built on Soei. Observation (Soei core) stays read-only, as required by P1. Any enforcement lives in a separate policy plane that consumes Soei verdicts, so it remains an explicit, auditable and reversible decision.

### 13.3 Harness metrics with real-data baselines

**Table 11.** Golden-harness metrics. Baselines are measured on our 45-session corpus (human in the loop); targets are proposals for granting autonomy.

| Metric | Definition | Pilot baseline | Proposed L3 target |
|---|---|---|---|

| Metric | Definition | Pilot baseline | Proposed L3 target |
|---|---|---|---|
| **UHR** Unverified hand-back rate | Code-changing episodes ending with no proof after the last edit | 23.4% [17.1, 35.7] | ≤ 2% |
| **CP** Claim precision | Completed claims that are Verified (1 − unproved completion rate) | 42.2% (136/322) | ≥ 95% |
| **VCR** Verified completion rate | Tasks reaching Verified with a passing oracle | needs oracle outcomes | task-class specific |
| **PR** Phantom rate | Mutations not acknowledged | 1.1% (shown correctly by Soei) | 0 rendered as applied |
| **LC** Lineage coverage | Cross-session reuse explained by H2 lineage | 24/24 pairs (path level) | digest level, 100% |
| **SLR** Secret leak rate | Credential matches in persisted or exported records | raw 1,326 → Soei 0 | 0 |
| **TTA** Time to awareness | Time for a supervisor to answer "what is done and verified?" | 195 long-idle returns | user study (RQ5) |
| **RD** Replay determinism | Identical projections on replay | ✓ | ✓ always |

### 13.4 Autonomy levels and evidence gates

**Table 12.** Proposed autonomy levels for coding agents and the Soei evidence each requires before the level is granted.

| Level | Agent may… | Human role | Required Soei evidence (gate) |
|---|---|---|---|
| L0 Suggest | propose code only | applies every change | none (capture optional) |
| L1 Assisted edit | edit files in a workspace | approves each episode | ack-correct capture; UHR reported to the human at hand-back |
| L2 Verified edit | edit and hand back without per-step approval | reviews Unverified/Contested only | every claim Verified or explicitly flagged; UHR ≤ target; SLR = 0 |
| L3 Multi-agent autonomy | run sessions and sub-agents that build on each other | samples audits | digest lineage for every hand-off; CP ≥ target on the golden set; contested-claim alerts |
| L4 Unattended delivery | merge and deploy within policy | sets policy, reviews audit | signed provenance attestation per artifact; oracle-verified VCR; replayable ledger; automatic revocation on regression |

The gates make autonomy something an agent earns and can lose. An agent is granted $L_k$ for a task class when its golden-harness scorecard meets the gate. It is automatically demoted when replayed regression runs or production evidence push a metric past threshold, for example a model upgrade that raises UHR.

### 13.5 Alignment with governance frameworks

**Table 13.** How Soei evidence supports common AI-governance obligations (indicative mapping, not legal advice).

| Framework | Obligation (summary) | Soei support |
|---|---|---|
| NIST AI RMF 1.0 [13] | GOVERN, MAP, MEASURE, MANAGE functions; measurable, documented risk management | Harness metrics (MEASURE), autonomy gates (MANAGE), ledger-backed records (GOVERN) |
| ISO/IEC 42001:2023 [14] | AI management system: operational control, monitoring, internal audit, continual improvement | Tamper-evident monitoring record; replay for audit; regression alarms for improvement |
| EU AI Act, Art. 12 [15] | Automatic recording of events (logging) over the lifetime of high-risk systems | Append-only, hash-chained, policy-tagged event log with manifests |
| EU AI Act, Art. 14 [15] | Effective human oversight, including understanding outputs and deciding not to use them | Claim vs proof display, Contested alerts, attention ranking, explainable support links |
| Supply-chain provenance (SLSA / in-toto) [16, 17] | Verifiable provenance for build artifacts | Planned: export Soei lineage and proof as attestations for agent-produced commits |

**Why it cannot be retrofitted.** Governance evidence has to exist *at the time of the action*. History that was not captured, was captured unacknowledged, was bounded away, or was polluted with secrets cannot be reconstructed for an audit later. Our measurements show all four failure modes in real logs (§10.4–10.6). The substrate has to be in place before autonomy is granted, not added afterwards.

## 14 Scope of improvements

The prototype shows the core is feasible and cheap. The improvements below are what remain between it and a production golden harness, ranked by governance impact. Each lists its current state, the change, and why it is required.

**Table 14.** Prioritised improvement backlog.

| # | Improvement | Current state (prototype) | Change | Why required | Phase |
|---|---|---|---|---|---|
| 1 | **Proof outcomes** | Detects that a check ran, not whether it passed | Capture exit codes, test reports (JUnit/pytest), HTTP status, UI assertions; enable the *Contested* state | A failed check is currently counted as proof; L2+ gates need pass/fail | 1 |
| 2 | **Digest-level lineage** | Path-hash identity | Content digests (BLAKE3/Git blob) at every write and read | Path reuse after overwrite can create false hand-offs; L3 accountability needs exact versions | 1 |
| 3 | **Learned proof classifier** | Rules v4, audited precision 83.3% / 96.7% | Train on audited labels; per-class calibration; abstain to "Unknown"; evaluate trajectory verifiers such as SWE-Critix [31] as drop-in classifiers | Gate decisions need ≥95% precision with calibrated uncertainty | 2 |
| 4 | **Multi-agent adapters** | VS Code Copilot agent only | Codex, Claude Code, Cursor, MCP servers, CI systems, with conformance corpora | Fleets mix vendors; one-vendor governance has blind spots | 1–2 |
| 5 | **Golden task corpus + oracles** | None | Curated, versioned tasks from real work with hidden oracle tests (incl. post-silicon validation tasks) | Certification needs ground truth, not self-reported success | 2 |
| 6 | **Policy plane** | Not present (core is read-only by design) | Separate, opt-in plane: advisory → enforcing merge/deploy gates from Soei verdicts; signed policy files | Turns evidence into governance; keeps observation and control separable | 3 |
| 7 | **Provenance attestations** | Not present | Emit SLSA/in-toto-style statements linking commits to lineage, proof and agent identity | Downstream consumers and auditors can verify agent-made changes | 3 |
| 8 | **Secret defence in depth** | Pattern-based scan, gate before persistence | Entropy and context detectors, provider-side redaction hooks, live alerts on exposure | 1,326 raw records exposed in 16/45 sessions | 1 |
| 9 | **Causal ordering** | Wall clock + source offsets | Hybrid logical clocks across sources and machines | Remote and sub-agent runs break wall-clock order | 2 |
| 10 | **Plan-aware attribution and drift** | Not evaluated | Map artifacts to plan-file components; calibrated attribution; drift metrics (Eq. 2) | Supervisors reason in components and deliverables, not files | 2 |
| 11 | **Differential replay** | Determinism verified | Replay identical tasks across model/scaffold versions; regression alarms | Model upgrades silently change behaviour; autonomy must be revocable | 2 |
| 12 | **Supervisor UX** | No UI yet | Graph/role/timeline projections with Verified/Unverified/Contested, attention queue, "why" links | Oversight only works if evidence reaches the human quickly (RQ5) | 3 |
| 13 | **Team federation** | Single machine | Opt-in aggregation of metrics (not content) across a team; privacy-preserving statistics | Governance and benchmarking happen at team and org level | 4 |

| # | Improvement | Current state (prototype) | Change | Why required | Phase |
|---|---|---|---|---|---|
| 14 | **Scale and retention** | 15 MB per 5 developer-months | Segment compaction, retention policy, SQLite/columnar storage, streaming fold | Fleet-scale autonomy multiplies event volume | 4 |

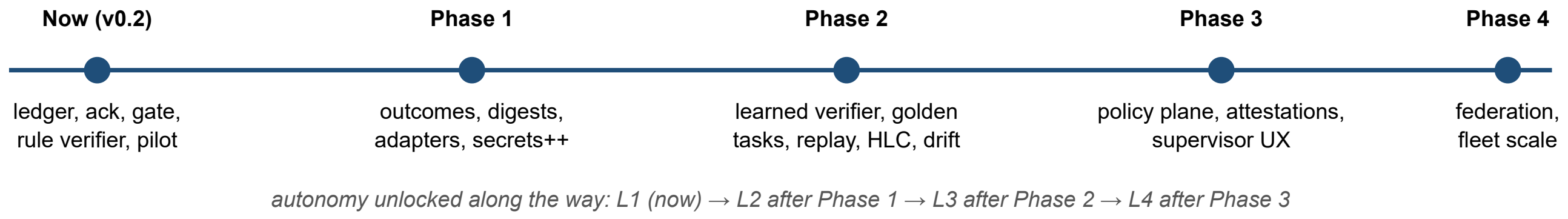

**Figure 18.** Improvement roadmap and the autonomy level each phase is intended to unlock.

## 15 Conclusion

Local, project-level observability of coding agents is feasible, and developers want it. Our study of 45 real sessions shows that visibility on its own is not enough. About a quarter of code-changing interactions end unverified. More than half of agent-completed, edit-backed to-dos have no proof. Timing-based hand-off detection sees a quarter of real collaboration. Bounded history drops the long sessions that hold half the evidence. Raw agent logs routinely contain credentials. Soei responds with one principle at each layer: acknowledge before showing, persist evidence before inferring, scrub before storing, prove before calling something done, and derive every view from one tamper-evident record. The prototype shows that this costs about 20% of the raw log size and fractions of a second on real data. The same substrate is what a golden harness and the governance of autonomous agents require: autonomy that is granted on measured evidence, verified per claim, attributable per artifact, auditable after the fact, and revoked automatically when the evidence degrades.

## Appendix A  Soei Event Schema examples

```
{ "specversion": "soei/0.1", "id": "01JC9Z6Q8W3T0E6M8X2D5V4K7A", "type": "artifact.generated",
  "source": "adapter:copilot-vscode/0.56#transcripts/<session>.jsonl", "source_offset": 1842,
  "hlc": { "wall": "2026-10-07T11:43:48.454Z", "logical": 0, "node": "src:copilot:<s>" },
  "evidence_class": "observed",
  "ack": { "state": "acknowledged", "call": "c1f0...", "result_ts": "2026-10-07T11:43:48.920Z" },
  "subject": { "session": "9a3e...", "artifacts": ["path#5b1d..."], "code": true },
  "policy": { "profile": "default-local", "dropped": ["message", "arguments.content"], "hashed": ["path"] },
  "prev": "a052e782...", "hash": "7c19b0e4..." }

{ "specversion": "soei/0.1", "type": "proof", "evidence_class": "observed", "proof": "test",
  "ack": { "state": "acknowledged" }, "subject": { "session": "9a3e...", "activity": "4410..." } }

{ "specversion": "soei/0.1", "type": "task.update", "evidence_class": "reported",
  "ack": { "state": "acknowledged" }, "tasks": [ { "id": "t#e21c...", "status": "completed" } ] }
```

## Appendix B  Gap-to-requirement traceability

| Gap | Requirement | Soei component | Evidence / evaluation |
|---|---|---|---|
| G1/G2 | Single backend, precedence, namespaced hooks | L1–L2, L6 | Replay determinism ✓; RQ4 |
| G3 | Formal evidence, calibration | §6, L4 | Audited precision; RQ1 |
| G4 | Acknowledgement | Ack tracker | 1.1% phantom edits (§10.5) |
| G5 | Claim versus proof | L5 | 23.4% / 57.8% (§10.1); RQ3 |
| G6 | Graded lineage hand-offs | L4 lineage | Recall 25% for H0 (§10.3); RQ2 |
| G7 | Measured attribution | L4 attribution | RQ1 |
| G8 | Durable history | L2 ledger, checkpoints | 48.1% bytes beyond bound (§10.4) |
| G9 | Causal order | HLC, offsets | RQ4 |
| G10 | Security, privacy | Gate, token, Origin/Host | 1,326 → 0 (§10.6) |
| G11–G13 | Adapters, interop, attention | SDK, exporters, L4 | Windows prototype run; 195 idle returns |
| G14 | Evaluability | Harness | This paper |

## Appendix C  Proof classifier rules (v4, abridged)

| Class | Rule (applied to acknowledged actions after the last code edit) |
|---|---|
| static | Editor diagnostics; `py_compile`, `compileall`, ruff/flake8/pylint/mypy, `tsc`, eslint, `node --check`, JSON validation |
| build | `npm run build\|compile\|lint\|typecheck`, dotnet/msbuild/cmake/make/gradle/mvn/cargo/go build, `vsce package`, choco/nuget/npm pack, `pip wheel` |
| test | pytest, unittest, `npm test`, go/cargo/dotnet/mvn/gradle test, ctest; scripts named `test*/verify*/check*/smoke*` |
| exec | Interpreter runs the edited script or a script in the edited directory; `&`/`-File` runs an edited `.ps1`; notebook cell runs. Excludes invocations followed by sync/put/deploy/upload/scp/copy |
| smoke | `python -c` with import of project modules |
| probe | Invoke-WebRequest/RestMethod or curl to 127.0.0.1/localhost |
| ui | Browser page read, navigation, screenshot or scripted checks after an edit in the episode |

## Appendix D  Audit record

| Version | Label | Seed | Correct / n | Typical errors |
|---|---|---|---|---|
| v1 | unverified | 7 | 20 / 30 | build commands, quoted interpreter paths, import smoke checks, `_tmp_test` scripts |

| Version | Label | Seed | Correct / n | Typical errors |
|---|---|---|---|---|
| v3 | unverified | 41 | 22 / 30 | Playwright UI checks, localhost HTTP probes, `&`-invoked `.ps1`, `choco pack` |
| v4 | unverified | 59 | 25 / 30 | log-tail inspection, grep-based rename check, terminal output review, subagent review |
| v4 | verified | 59 | 29 / 30 | `scp` deploy read as execution |

## Appendix E Reproducibility

```
research/soei/prototype/
  soei_pilot.py        # adapter + SES + ack + privacy gate + hash-chained ledger + fold + all metrics (stdlib
only)
  audit_sample.py      # seeded blind audit sampler:  python audit_sample.py {unverified|verified} <seed>
  make_figures.py      # all data figures (matplotlib) + bias-corrected estimate
  results_frozen.json  # aggregate results cited in this paper
  audit_results.json   # audit outcomes per version
python soei_pilot.py --out results.json --ledger ledger.jsonl  &&  python make_figures.py
```

The corpus itself is not released: it is the authors' private working history and contains secrets (§10.6). The scripts run unchanged on any VS Code Copilot agent transcript directory. Other agents need an adapter conforming to §5.2.